\documentstyle[12pt]{article}
\begin{document}
%{\large 
%%%%%%%%%%%%%%%%%%%%%%%%%%%%%%%%%%%%%%%%%%%%%%%%%%%%%%%%%%%%%%%%
\title{{\bf Computations of Scattering Lengths in $nnpp$ System 
within Cluster Reduction Method for Yakubovsky 
Equations\thanks{Contribution to XV International Conference on Few-Body 
problem in Physics, Groningen, the Netherlands, 22--26 July 1997} 
}}
%%%%%%%%%%%%%%%%%%%%%%%%%%%%%%%%%%%%%%%%%%%%%%%%%%%%%%%%%%%%%%%%%
\author{S.L. Yakovlev\thanks{E-mail: {\tt yakovlev@snoopy.phys.spbu.ru}}
 \ I.N. Filikhin}
%%%%%%%%%%%%
\maketitle 
%%%%%%%%%%%
\date{}
%%%%%%
\begin{center}
{\small {\it Department of Mathematical and Computational 
Physics, St.Petersburg State University,\\ 
Ulyanovskaya Str. 1, 198904 St.Petersburg, Russia}} 
\end{center}
%%%%%%%%%%%%%%%%%%%%%%%%%%%%%%%%%%%%%%%%%%%%%%%%%%%%%%%%%%%%%%%%%%%%%%%%%%%%%%
\section{INTRODUCTION}
The elastic and rearrangement processes in the four nucleon system with two 
clusters in the initial and final states can be treated 
adequately in framework of Yakubovsky differential equations (YDE) approach. 
Unfortunately, a direct application of YDE to the scattering problem requires 
huge computer resources. It is why, new methods allowing a reduction of 
comlexity of YDE are of interests from the point of view of practical calculations. 
In papers \cite{1}, the authors proposed a method which reduces the YDE to 
the equations for the functions describing the relative motions of clusters. 
This method of cluster reduction (CRM) was successfully applied to 
calculations of $n$--$^3$H scattering in \cite{1}. It is worth to note that 
the exact four nucleon calculations of $n$--$^3$H scattering were performed 
on a personal computer what can characterize an efficiency of CRM.

In this report we give a sketch of CRM and our recent results of calculations 
of scattering lengths in the four nucleon system.
%%%%%%%%%%%%%%%%%%%%%%%%%%%%%%%%%%%%%%%%%%%%%%%%%%%%%%%%%%%%%%%%%%%%%%%%%%%%%%
\section{CRM FORMALISM}
In this section we present a brief review of CRM for the low energy 
scattering problem in the system of four neutral particles. 
The more comprehensive description can be found in the papers \cite{1}.
The generalization on the case of charged particles is straightforward 
following the results of \cite{2}.

The starting point is YDE approach. In framework of YDE the four body wave 
function should be decomposed into components $\Psi_{a_3 a_2}$ in one to one 
correspondence to the all chains of three cluster ($a_3$) and two cluster 
($a_2$) partitions. The components $\Psi_{a_3 a_2}$ obey the YDE \cite{3}
%%%%%%%%%%%%%%%%%%%%%%%%%%%%%%%%%%%%%%%%%%%%%%%
\begin{equation}
(H_0+V_{a_3}-E)\Psi_{a_3 a_2} + V_{a_3}\sum_{(c_3\ne a_3)\subset a_2}
\Psi_{c_3 a_2} = -V_{a_3}\sum_{d_2 \ne a_2}\sum_{(d_3\ne a_3)\subset 
a_2}\Psi_{d_3 d_2}
\label{1}
\end{equation}
%%%%%%%%%%%%%%%%%%%%%%%%%%%%%%%%%%%%%%%%%%%%%%%%
For the two clusters collisions the YDE admit a further reduction. Let 
$$
H_0=T_{a_2}+T^{a_2}
$$
be the separation of kinetic energy operator in the intrinsic $T_{a_2}$ 
with respect to the clusters of $a_2$ part and the kinetic energy $T^{a_2}$
 of the relative motion of $a_2$ clusters. The cluster reduction procedure 
consists in  expanding  of the components $\Psi_{a_3 a_2}$ along the basis 
of the solutions of Faddeev equations (FE) for subsystems of the partition $a_2$
%%%%%%%%%%%%%%%%%%%%%%%%%%%%%%%%%%%%%%%%%%%%%%%%%
$$
(T_{a_2} + V_{a_3})\psi^{a_3}_{a_2,k} +V_{a_3}\sum_{(c_3\ne a_3)\subset a_2}
\psi^{c_3}_{a_2,k} = \varepsilon^{k}_{a_2}\psi^{a_3}_{a_2,k}.
$$
%%%%%%%%%%%%%%%%%%%%%%%%%%%%%%%%%%%%%%%%%%%%%%%%%
The actual expansion has the form
%%%%%%%%%%%%%%%%%%%%%%%%%%%%%%%%%%%%%%%%%%%%%%%%%%
\begin{equation}
\Psi_{a_3 a_2} = \sum_{k\ge 0} \psi^{a_3}_{a_2,k}({\bf x}_{a_2})F^{k}_{a_2}(
{\bf z}_{a_2})
\label{2}
\end{equation}
%%%%%%%%%%%%%%%%%%%%%%%%%%%%%%%%%%%%%%%%%%%%%%%%%
Here the unknown amplitudes $F^{k}_{a_2}({\bf z}_{a_2})$ depend only on the 
relative position vector ${\bf z}_{a_2}$ between the clusters of the partition 
$a_2$. The basis of the solutions of FE is complete but not the orthogonal one
\cite{4} due to not Hermitness of FE. The biorthogonal basis is formed by the 
solutions of conjugated FE 
%%%%%%%%%%%%%%%%%%%%%%%%%%%%%%%%%%%%%%%%%%%%%%
$$
(T_{a_2} + V_{a_3})\phi^{a_3}_{a_2,k} + \sum_{(c_3\ne a_3)\subset a_2}
V_{c_3}\phi^{c_3}_{a_2,k} = \varepsilon^{k}_{a_2}\phi^{a_3}_{a_2,k}.
$$
%%%%%%%%%%%%%%%%%%%%%%%%%%%%%%%%%%%%%%%%%%%%%%%%%%%%%%%%%%%%%%%%%%%%%%%%%%
Introducing expansion (\ref{2}) into YDE (\ref{1}) and projecting onto the 
elements of biorthogonal basis $\{ \phi^{a_3}_{a_2,k}\} $ lead to the 
resulting reduced YDE \cite{1} for $F^{k}_{a_2}({\bf z}_{a_2})$:
%%%%%%%%%%%%%%%%%%%%%%%%%%%%%%%%%%%%%%%%%%%%%%%%%%%%%%%%%%%%%%%%%%%%%%%%%%%
\begin{equation}
(T^{a_2} - E + \varepsilon^{k}_{a_2})F^{k}_{a_2} = 
-\sum_{{a_3}\subset {a_2}}\langle \phi^{a_3}_{a_2,k}|V_{a_3}
\sum_{d_2\ne a_2}\sum_{(d_3\ne a_3)\subset a_2}\sum_{l\ge 0}
\psi^{d_3}_{d_2,l}F^{l}_{d_2}\rangle .
\label{3}
\end{equation}
%%%%%%%%%%%%%%%%%%%%%%%%%%%%%%%%%%%%%%%%%%%%%%%%%%%%%%%%%%%%%%%%%%%%%%%%%%
Here the brackets $\langle .|. \rangle $ mean the integration over 
${\bf x}_{a_2}$. The asymptotic boundary conditions for $F^{k}_{a_2}({\bf z}
_{a_2})$ have the following {\it two body} form as $|{\bf z}_{a_2}|
\rightarrow \infty $: 
%%%%%%%%%%%%%%%%%%%%%%%%%%%%%%%%%%%%%%%%%%%%%%%%%%%%%%%%%%%%%%%%%%%%%%%%%
\begin{equation}
F^{k}_{a_2}({\bf z}_{a_2}) \sim \delta_{k0}[ 
\delta_{{a_2}{b_2}}\exp i ({\bf p}_{a_2},{\bf z}_{a_2}) +
{\cal A}_{{a_2}{b_2}}\frac{\exp i \sqrt{E-\varepsilon^{0}_{a_2}}|
{\bf z}_{a_2}|}{|{\bf z}_{a_2}|} ] ,
\label{4}
\end{equation}
%%%%%%%%%%%%%%%%%%%%%%%%%%%%%%%%%%%%%%%%%%%%%%%%%%%%%%%%%%%%%%%%%%%%%%%%%
where it is implied that the subsystems of partitions $a_2$ have only one 
bound state with $\varepsilon^{0}_{a_2}$ being the respective binding energy.
The subscript $b_2$ corresponds to the initial state and ${\bf p}_{a_2}$ is 
the conjugated to ${\bf z}_{a_2}$ momentum.
%%%%%%%%%%%%%%%%%%%%%
%%%%%%%%%%%%%%%%%%%%%%%%%%%%%%%%%%%%%%%%%%%%%%%%%%%%%%%%%%%%%%%%%%%%%%%%%%%%
\section{RESULTS of CALCULATIONS of SCATTERING LENGTHS in FOUR NUCLEONS SYSTEM}
%%%%%%%%%%%%%%%%%%%%%%%%%%%%%%%%%%%%%%%%%%%%%%%%%%%%%%%%%%%%%%%%%%%%%%%%%%%%
%%%%%%%%%%%%%%%%%%%%%
In this section we present the results of calculations of channels scattering 
lengths in $N-NNN$ system (without Coulomb interactions) for the states with 
the total isospin $T=0$, and results for scattering lengths in $d$--$d$, 
$n$--$^3$He and $p$--$^3$H systems. The MT I-III potential model was used to 
describe the $N$--$N$ interaction. 

In the first case the total spin $S$ and isospin $T$ are the integrals of 
motion, so that the scattering lengths $A_{ST}$ can be introduced. The results 
of our calculations together with data of other authors are collected in the 
Table 1.
\vskip 1cm
%\newpage
\begin{center} 
Table 1. Results of calculations of $A_{ST}$ for $S=0,1$, $T=0$ states.\\ 
\vskip 1cm
\begin{tabular}{|c|c|c|}
\hline \hline 
\parbox{4cm}{\center Refs. }&
\parbox{4cm}{\center $A_{10}$ \ fm }& 
\parbox{4cm}{\center $A_{00}$ \ fm } \\
                     &                     &            \\
\hline
This work            & 2.8                 & 14.7      \\
\hline
\cite{5}             & 3.013               & 12.317     \\ 
\hline
\cite{6}             & 3.09                & 14.95       \\ 
\hline
\cite{7}             & 2.9                 & 8.1          \\ 
\hline            
\cite{8}             & 3.25                & 14.75       \\
\hline \hline 
\end{tabular}  
\end{center}
\vskip 2cm 

The modified version of equations (\ref{3}) followed from equations \cite{2} 
was used for computations of scattering lengths in the $nnpp$ system taking 
into account the Coulomb interaction between protons. The   $d$--~$d$ 
and $n$--$^3$He, $p$--$^3$H scattering lengths 
corresponding to the values of total 
spin $S=0,2$ for $d$--$d$ scattering and $S=0,1$ for $n$--$^3$He and $p$--$^3$H
 scattering are collected in Tables 2,3 and 4. 
%%%%%%%%%%%%%%%%%%%%%%%%%%%%%%%%%%%%%%%%%%%%%%%%%%%%%%%%
\newpage 
\vskip 1cm
\begin{center}
Table 2. $d$--$d$ scattering lengths for $S=0$ and $S=2$ states. \\
\vskip 1cm 
\begin{tabular}{|c|c|c|}
\hline \hline
\parbox{4cm}{\center Refs.} &
\parbox{4cm}{\center $A_{d-d} (S=0)$ fm} &
\parbox{4cm}{\center $A_{d-d} (S=2)$ fm} \\ 
           &                     &                 \\ 
\hline 
This work  &  10.2 - 0.2 $i $ &    7.5          \\ 
\hline \hline   
\end{tabular}
\end{center}
%%%%%%%%%%%%%%%%%%%%%%%%%%%%%%%%%%%%%%%%%%%%%%%%%%%%%%%%%%%%%%%%%%%
%\newpage 
\vskip 1cm
\begin{center}
Table 3. $p$--$^3$H and $n$--$^3$He scattering lengths for $S=0$ state.
\vskip 1cm 
\begin{tabular}{|c|c|c|}
\hline \hline 
\parbox{3.5cm}{\center Refs. } & 
\parbox{3.5cm}{\center $A_{p-{^3}H}$ fm. } & 
\parbox{5cm}{\center $A_{n-{^3}He}$ fm. } \\ 
                   &                        &                   \\ 
\hline 
This work          & -22.6         &  7.5 - 4.2 $i$          \\ 
\hline 
\cite{9}           & -21           &  7.16 - 3.9 $i$           \\ 
\hline 
\cite{7}           & 4.2           &  6.05 - 0.72 $i$           \\ 
Exp. \cite{10}     &  -            & 6.53 ($\pm $ 0.32) - 4.445 ($\pm $ 
0.003) $i$ \\  
\hline \hline 
\end{tabular} 
\end{center}
%%%%%%%%%%%%%%%%%%%%%%%%%%%%%%%%%%%%%%%%%%%%%%%%%%%%%%%%%%%%%%%%%%%%%%%%%%%%
\vskip 1cm 
\begin{center}
Table 4. $p$-$^3$H and $n$-$^3$He scattering lengths for $S=1$ state.
\vskip 1cm 
\begin{tabular}{|c|c|c|}
\hline \hline 
\parbox{4cm}{\center Refs.} & 
\parbox{4cm}{\center $A_{p-{^3}H}$ fm } & 
\parbox{4cm}{\center $A_{n-{^3}He}$ fm }  \\ 
                       &                    &                   \\ 
\hline 
This work              & 0.5        & 2.9 - 0.0 $i$ \\ 
\hline 
\cite{7}               &  -         & 4.25 + 0.005 $i$\\ 
\hline \hline
\end{tabular}
\end{center}
%%%%%%%%%%%%%%%%%%%%%%%%%%%%%%%%%%%%%%%%%%%%%%%%%%%%%%%%%%%%%%%%%%%%%%%%%%%%
\vskip 1cm 

The modified equations (\ref{3}) were used to calculate the $0^{+}$ resonance 
state of $^4$He. The results of calculations are presented in the Table 5.
\newpage 
\vskip 1cm 
\begin{center}
Table 5. Results of calculations for the energy $E=E_{r} + i \frac{\Gamma}
{2} $ \\ of $0^{+}$ resonance state of $^4$He.
\vskip 1cm 
\begin{tabular}{|c|c|c|}
\hline \hline 
\parbox{4cm}{\center Refs.} & 
\parbox{4cm}{\center $E_{r}$ MeV.} & 
\parbox{4cm}{\center $\Gamma $ MeV. } \\ 
    &   &  \\ 
\hline
This work     & 0.15     & 0.3   \\ 
\hline 
\cite{9}      & 0.12     & 0.26   \\ 
\hline 
Exp. \cite{11} & 0.25    & 0.3     \\
\hline  \hline
\end{tabular} 
\end{center} 
%%%%%%%%%%%%%%%%%%%%%%%%%%%%%%%%%%%%%%%%%%%%%%%%%%%%%%%%%%%%%%%%%%%%%%%%%%%
\section{ACKNOWLEDGEMENTS}
This work was partially supported by the Russian Foundation for Basic
Research (Project No. 96-0217031). One of the authors (S.L.Y.) is grateful 
to Prof. L.P. Kok Institute for Theoretical Physics University of Groningen 
for the kind invitation  to visit Groningen University during the period of 
Few Body XV Conference. 
%%%%%%%%%%%%%%%%%%%%%%%%%%%%%%%%%%%%%%%%%%%%%%%%%%%%%%%%%%%%%%%%%%%%%%%%%%%%
%\newpage

%}
\end{document}